# An Investigation On Neck Extensions For Single and Multi-Degree Of Freedom Acoustic Helmholtz Resonators

Abhishek Gautam[1], Alper Celik[2] and Mahdi Azarpeyvand[3]
*Faculty of Engineering, University of Bristol, Bristol, United Kingdom*

The effect of neck extensions in single and multi-degree of freedom Helmholtz resonator based acoustic liners is studied both experimentally and numerically and the resulting transmission coefficient and resonance frequencies are examined. It has been shown that a single degree of freedom liner with increasing neck extension lengths leads to the resonance frequencies being pushed to lower frequency values, however, this shift to lower frequencies is not linear with increasing length. A study on including neck extensions for the primary and/or secondary neck within a double degree of freedom liner is also presented. It is shown that both neck extension concepts lead to an increase in bandwidth of sound absorption by a double degree of freedom Helmholtz resonator.

## I. Nomenclature

| | | |
|---|---|---|
| *DoF* | = | Degree of freedom |
| *TL* | = | Transmission loss |
| *L* | = | Cavity height of baseline single degree of freedom liner |
| *L1* | = | Height of the first cavity for the double degree of freedom liner |
| *L2* | = | Height of the second cavity for the double degree of freedom liner |
| T | = | Neck length of the single degree of freedom liner |
| *T1* | = | Neck length of the first neck for the double degree of freedom liner |
| *T2* | = | Neck length of the second neck for the double degree of freedom liner |
| *H* | = | Neck extension length |
| *V1* | = | Volume of first cavity for the double degree of freedom liner |
| *V2* | = | Volume of second cavity for the double degree of freedom liner |
| *PML* | = | Perfectly matched layer |
| Φ | = | Neck extension ratio |
| *PLA* | = | Polylactic acid |

## II. Introduction

Aero engine fan noise has been growing in dominance when it comes to the acoustic field radiated from an aircraft engine, especially for stages of the flight envelope dominated by high thrust operations like take off. With the aircraft industry heading towards more efficient higher bypass ratio engines, fan noise can only be assumed to grow in dominance.

Attenuating sound at its source is one of the ways to effectively absorb the low frequency content of the radiated sound field. Acoustic liners treatments are therefore applied on the inside wall of the engine nacelle, both at the inlet and in the bypass ducts. One of the most common liner treatments are based on Helmholtz resonators which are devices based on the principle of Helmholtz resonance where the resonance frequency depends on geometric parameters such as the cavity volume and cavity neck dimensions. The expression of Helmholtz resonance is as follows: $f_r = \left(\frac{c_0}{2\pi}\right)\sqrt{A_n/\{V_c(l_n + \delta_n)\}}$. Here $c_0$ is the speed of sound, $A_n$ is the cross sectional area of the neck, $V_c$ is the volume of the resonator, $l_n$ is the length of the neck and $\delta_n$ is an end correction factor, which is used to take into account the generation of higher order modes due to discontinuities in the geometry [1,2].

---

The effect of changing geometrical parameters like neck cross sectional area, location, shape and length, on the resonance frequency of a Helmholtz resonator with both a circular and square cavity was studied by Ingard [1]. End corrections from Ingard's investigation were used by Chanaud [2] to investigate the effect of changing the geometry of the orifice and cavity on the resonance frequency. The effect of changing width and depth of the cavity for a variable and fixed volume was also studied. The study reported that for a fixed cavity volume and orifice size, the shape of the orifice did not have a significant effect on the resonance frequency of the resonator, however, the position of the orifice affected the resonance frequency significantly.

Effect of changing cavity volumes and neck locations has also been studied by Selamet et al [3-5]. They investigated how different cavity length to diameter ratios would affect the transmission loss and resonance frequency. Since, the resonance frequency of a Helmholtz resonator is inversely proportional to the volume of the cavity, the challenge lies in reducing the resonance frequency without increasing the volume or reducing the volume to make the liner more compact without it leading to an increase in resonance frequency.

The effect of changing neck geometry in terms of cross-sectional area has been studied extensively in the past [6-9], however, the influence of extending the neck into the Helmholtz resonator cavity has not been investigated in detail. Therefore, the work in this paper will be focussed on the following:

(1) Experimental and numerical analysis of a single degree of freedom Helmholtz resonator with extended neck.

(2) Extending the study of neck extensions from a single degree of freedom resonator to a multi-degree of freedom resonator.

## III. Experimental Setup

Experiments are carried out at the University of Bristol Grazing Flow Impedance Tube Facility. The schematic of the facility is presented in Figure 1. The Impedance Tube is a 50.4mm x 50.4mm square cross-section tube, 4000 mm in length with a 3000 mm long diffusing section to reduce air velocity and minimise acoustic reflections back into the test section. Flow speeds of up to Mach 0.3 can be achieved within the test section using a 15kWh centrifugal fan. Sound pressure levels of up to 130dB were achieved in the test section via two BMS 4592ND compression drivers and microphone data was obtained via GRAS 40PL microphones. Data acquisition was achieved with a National Instruments PXIe-1082 data acquisition system. MatlabR2016a was used to interface between the data acquisition system and the signal generator to run the data acquisition code.

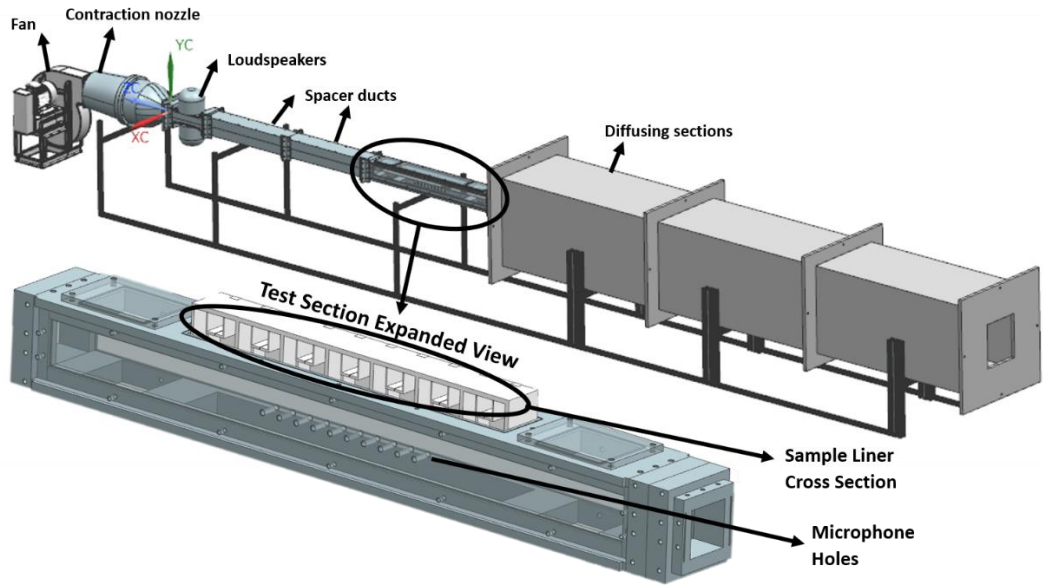

**Figure 1. Grazing Flow Impedance Tube at the University of Bristol**

For experimental analysis, a simple one degree of freedom resonator tuned to 670Hz ($f_0$) was additive manufactured alongside nine single degree of freedom liners cases with extended necks where the neck extension is a percentage of the baseline resonator cavity height, L. Schematics of the two types of liner cavities can be seen in Figure 2.

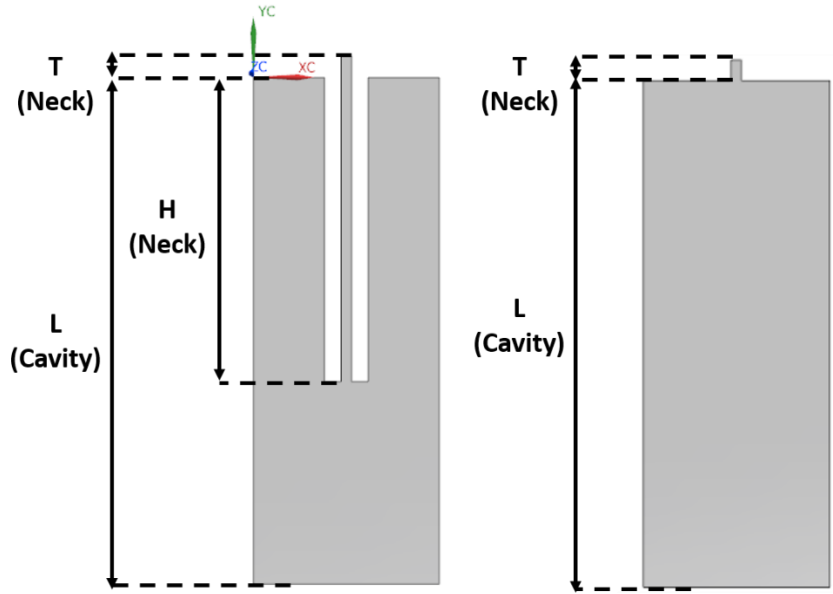

**Figure 2. (a) Extended neck liner cavity schematics; (b) Baseline Single Degree of Freedom Liner Schematics**

T and L are the neck length and cavity height respectively, of the designed Helmholtz resonator options. H is the length by which the baseline neck is extended. The neck extension for the liner options was altered by changing the extension ratio given by Φ where Φ is defined as the ratio of H/L and is swept from 0.1 – 0.9 for the experimental study. A Φ of 0.2 means that the neck extension, H, is 20 percent of the baseline single degree of freedom cavity height, L. Figure 3 shows an additive manufactured extended neck liner sample, with a Φ value of 0.4 and Figure 4 shows CAD geometry for a selection of extended neck liners designed.

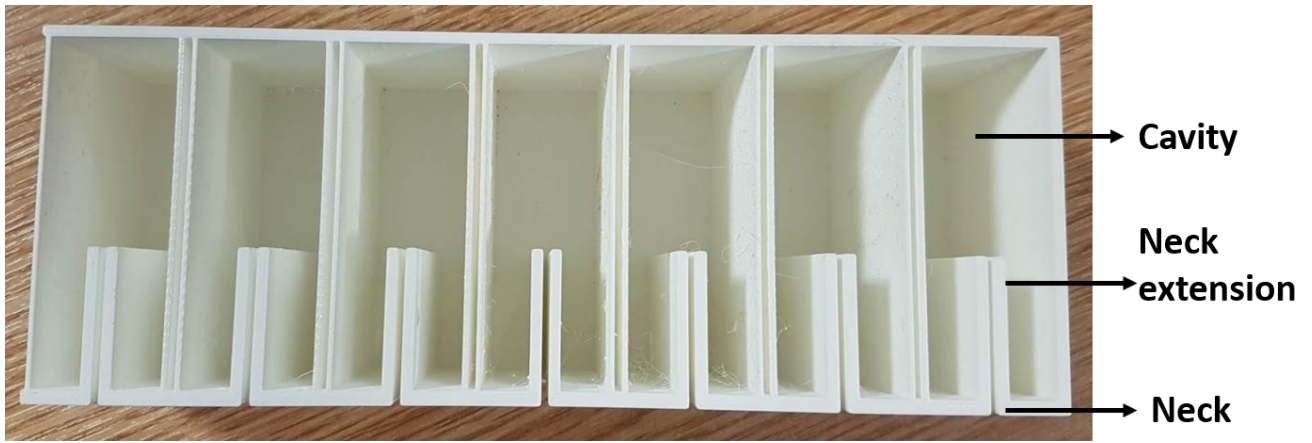

**Figure 3. Additive manufactured double degree of freedom liner option**

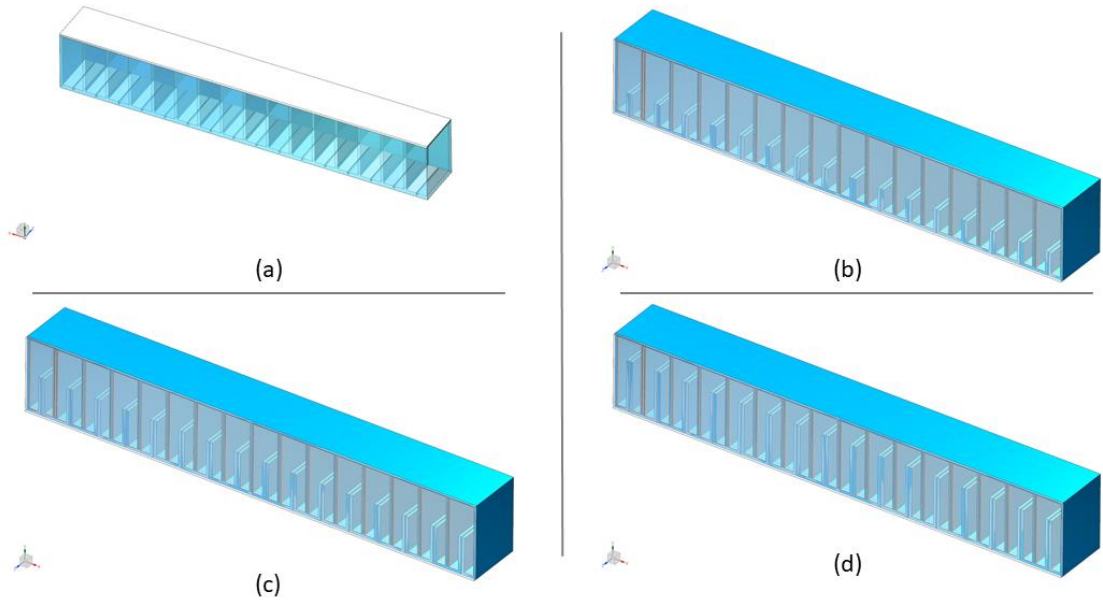

**Figure 4. CAD models for a selection of the liner concepts tested numerically and experimentally. (a) Baseline single degree of freedom liner; (b) Liner with Φ = 0.3; (c) Liner with Φ = 0.5; (d) Liner with Φ = 0.7**

## IV. Numerical Setup

Numerical simulations were carried out on COMSOL Multiphysics to obtain the transmission loss for the acoustic liners. A free triangular mesh was used for the impedance tube domain and a mapped mesh type was used for the acoustic box and acoustic termination. A Perfectly Matched Layer (PML) was used on either side of the impedance tube to prevent any reflections from affecting the results. The maximum element size for the mesh was chosen to be 2mm which is smaller than 1/6th of the wavelength of the highest frequency of interest i.e. 3000Hz. The minimum element size chosen was 0.12mm. The maximum element growth rate was 1.2 with a Curvature factor of 0.3 and Resolution of narrow regions of 3. The mesh quality can be seen in Figure 5. A background pressure field is applied to create a travelling plane wave. Transmission loss peaks obtained from steady state simulations were used to guide the transient simulations which were only computed at the peak frequencies to save computational time.

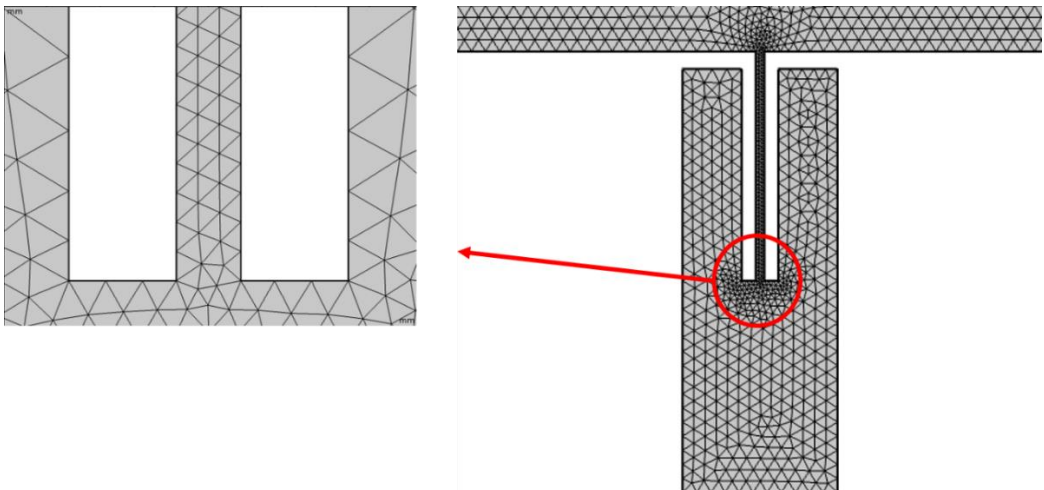

**Figure 5. Free triangular mesh used for Comsol simulations**

## V. Results

Liner options with Φ values ranging from 0.1 to 0.9 were numerically analysed on Comsol and the transmission loss calculated. It can be seen in Figure 6, that increasing the neck length into the cavity of the resonator leads to a reduction in resonance frequency of the liner cavity compared to a baseline single degree of freedom liner with a similar volume. Although the resonance frequency reduces with increasing neck length, the bandwidth of sound attenuation reduces with neck length as well i.e. the longer the neck, the narrower the band of frequencies attenuated.

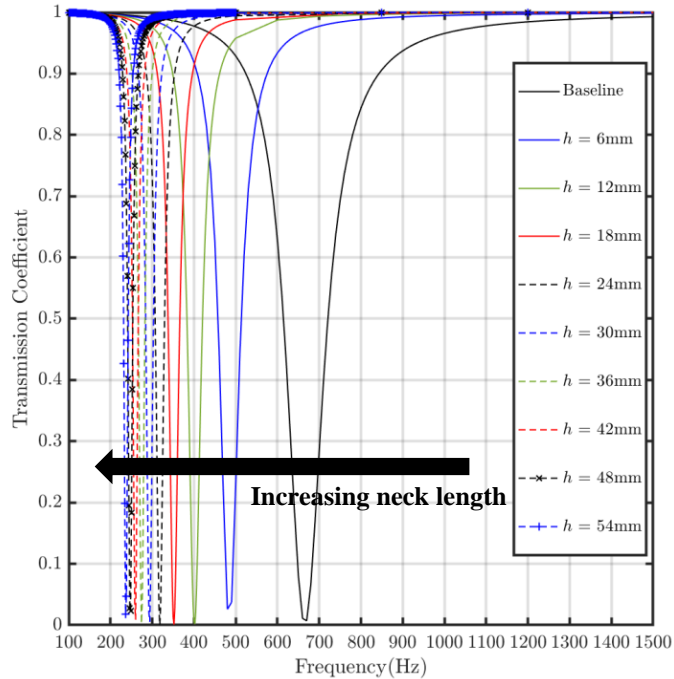

**Figure 6. Transmission Coefficient for resonators with Φ values from 0.1 – 0.9**

As seen in Figure 7, it was also observed that the reduction in resonance frequency with increase in neck length does not follow a linear behavior but decays rapidly up to a Φ value of 0.3 after which the performance of the resonator starts to stall and the frequency reduction is not as rapid. This may be an indication that there might be an optimum neck extension ratio where the trade-off between the bandwidth of sound attenuation and the frequencies attenuated is most favorable for low frequency applications. Further analysis work will be carried out to investigate this non-linear behavior of extending the neck into the cavity.

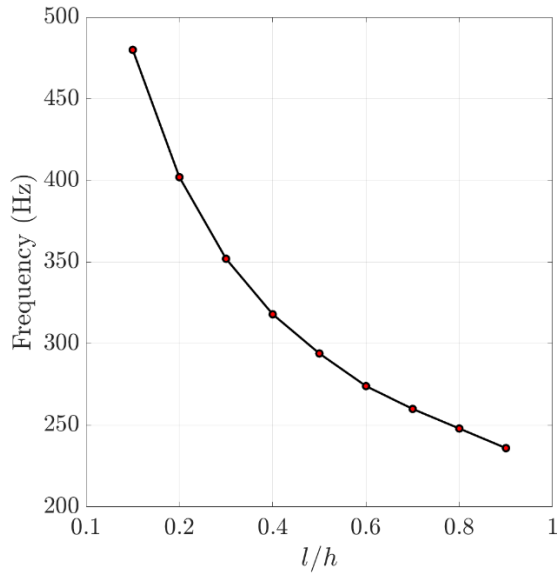

**Figure 7. Comparison of resonance frequency with increasing neck extension ratio**

Single degree of freedom liners with Φ values from 0.1 – 0.9, were additive manufactured in a 16-chamber configuration and the transmission loss induced by each sample was analysed in the Grazing Flow Impedance Tube. Figure 8 shows the experimental results of the transmission loss induced by the 16-chamber liner compared to numerical simulation results. The experimental and numerical results agree well for the 16-chamber liner configuration. There is a slight deviation in the transmission loss peak between experimental and numerical results for the single chamber configuration. This can be attributed to the fact that there is a finite input power to the speakers. The Comsol simulations assume a perfect sound hard boundary whereas in the case of experiments, the liners being additive manufactured using PLA, might not be completely sound hard which may be another cause for the deviation. Experimental studies will also be performed on single chamber liner configurations.

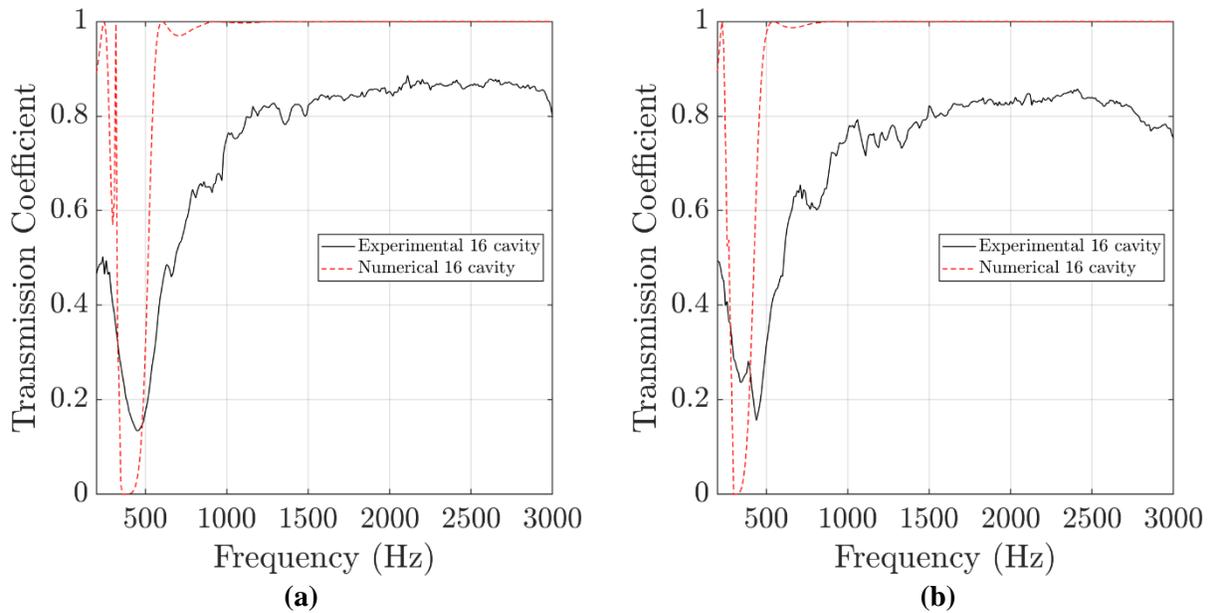

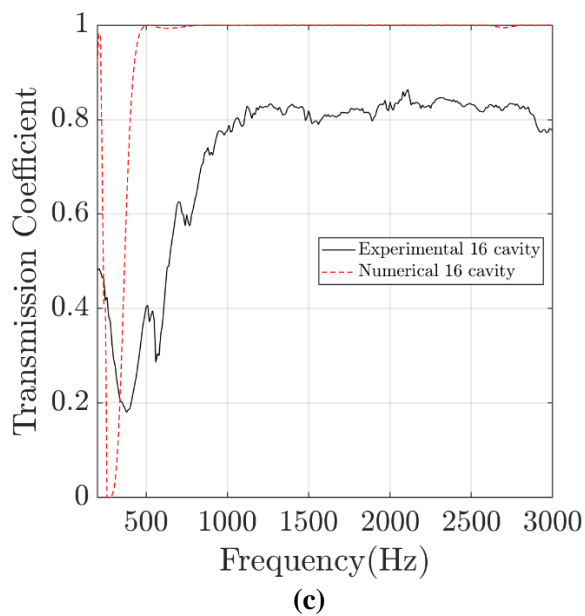

**(c)**

**Figure 8. Experimental and numerical simulation results for a selection of extended neck liner options in the 16-chamber configuration. (a) Φ = 0.3; (b) Φ = 0.5; (c) Φ = 0.7**

Study on the neck extensions for a single degree of freedom liner will also be extended to double degree of freedom liner cases. A double degree of freedom liner is one which has a pair of cavities and necks in a Neck-Cavity-Neck-Cavity configuration. Double degree of freedom liners have two resonance peaks, due to the presence of two cavities inside the resonating chamber. Compared to a baseline single degree of freedom liner, when the internal volume of the cavities for a 2DoF liner is altered, the secondary resonance frequency is substantially affected. As the volume of the first cavity increases, the difference between the first and second peak frequencies reduces up to a minimum after which it increases again. This might imply that there may be an optimal internal cavity volume ratio for the widest bandwidth of sound absorption.

Two different concepts of neck extensions were analysed for a double degree of freedom liner case as shown in Figure 9. In the first concept, the neck length for the middle septum was increased whereas in the second concept the primary neck was extended into the first cavity, to investigate the effect of changing neck lengths in a 2DoF liner, with changing internal cavity volume ratios.

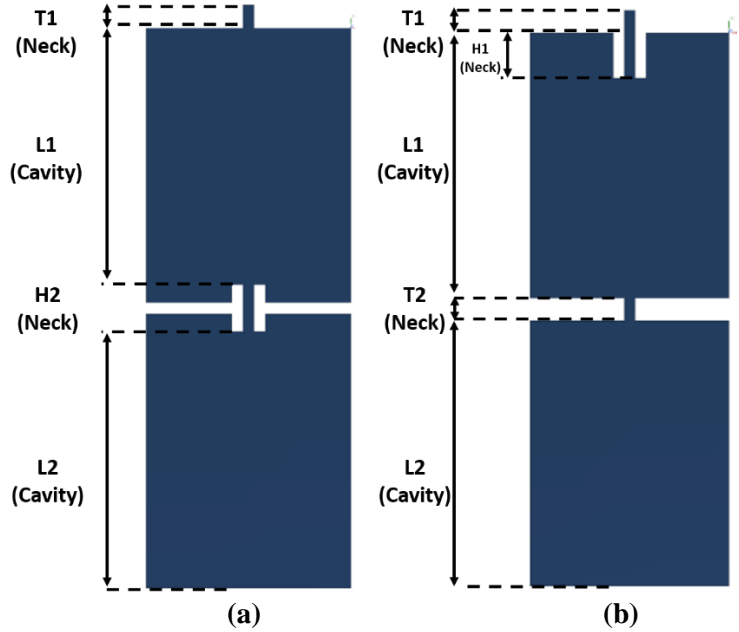

**Figure 9. 2DoF neck extension concepts; (a) Middle septum neck extension; (b) Primary neck extension into first cavity**

Figures 10 and 11 show the effect of neck extensions in a 2DoF liner. In the case of the middle septum extension, there is a substantial decrease in the difference between the first and second resonance frequencies for higher volume ratios whereas for the primary neck extension case, a similar decrease in the resonance frequency difference exists albeit not as substantial as the middle septum extension. However, both concepts exhibit an interesting behavior whereby one of them is efficient at higher volume ratios and the other one at lower volume ratios. Further investigation will be carried out around this behavior to find out if a hybrid between the two concepts can be used to manifest the performance of both concepts and improve the bandwidth of sound absorption via double degree of freedom Helmholtz resonator based acoustic liners.

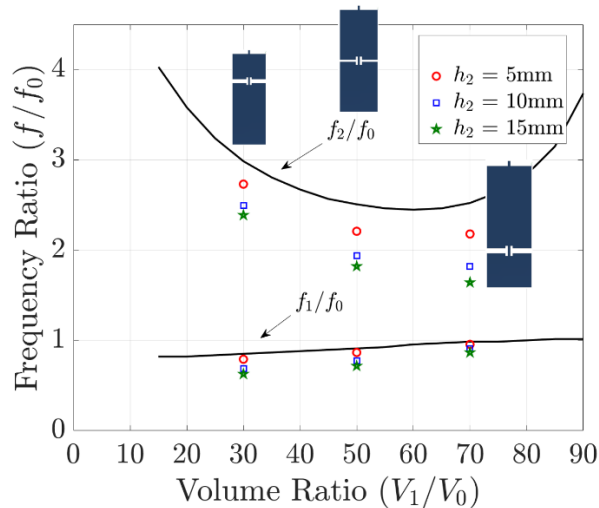

**Figure 10. Secondary neck length increase in a 2DoF Helmholtz Resonator**

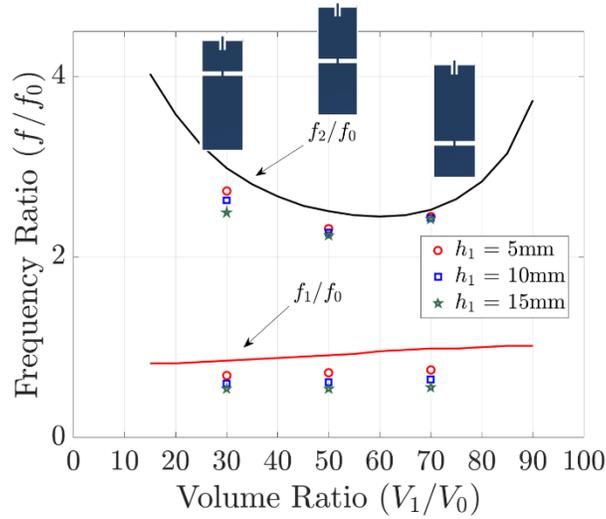

**Figure 11. Primary neck length increase in a 2DoF Helmholtz resonator**

## VI. Conclusion

The effect of extending the neck of a Helmholtz resonator into its cavity has been studied both experimentally and numerically in this paper. Extended neck resonators can be the sound attenuation devices for applications where there are space constraints, as it is shown that extending the neck reduces the resonance frequency of the device without any change to the overall volume of the resonator body. The nonlinear behavior of the reduction in resonance frequency with increasing neck length is shown and will be further investigated.

The neck length extension study for a single degree of freedom liner case has been extended to double degree of freedom liner concepts as well, where the middle septum and primary necks have been increased and their effects on the resonance frequencies are studied along with changing internal volume ratios for the 2DoF chambers. Both concepts show a promising nature of increasing the bandwidth of sound absorption by 2DoF liners by decreasing the difference between the primary and secondary resonance peaks with increasing neck lengths.

## VII. Future Work

For further studies, the authors will commit to further analysis of experimental data that has already been collected, which will include:
1. Studying the reason behind the nonlinear behavior of reduction in resonance frequency with increasing neck length.
2. Investigating the increasing bandwidth of sound absorption with the inclusion of neck extensions within 2DoF liners.
3. Modifying the double degree of freedom resonator geometry based on existing experimental and numerical results on neck extensions to further improve the bandwidth of sound absorption.